\documentclass{article}
\usepackage{amssymb}
\def\bv{{\bf v}}
\def\be{{\bf e}}
\def\bp{{\bf p}}
\def\bz{{\bf z}}
\def\bx{{\bf x}}
\def\by{{\bf y}}
\author{Cihan Tepedelenlioglu\\Department of ECEE \\ Arizona State University \\ cihan@asu.edu}
\begin{document}
\title{A Large-Deviation Upperbound on Directed Last-Passage Percolation Growth Rate Based on Entropy of Direction Vector}
\maketitle
\abstract{This short note provides a large-deviation-based upper bound on the growth rate of directed last passage percolation (LPP)  using the entropy of the normalized direction vector.}
\section{Growth Rate in LPP}
Consider the $d$-dimensional integer lattice $\mathbb{Z}^d$. For each point ${\bf z} \in \mathbb{Z}^d$ we associate a random variable $X({\bf z})$, with finite expectation $\mu := E[X({\bf z})]$. The stochastic process $X({\bf z})$ is i.i.d. across ${\bf z}$.

A directed path from the origin to ${\bf z}$ is a finite sequence of elements of
$\mathbb{Z}^d$ such that the difference of two consecutive elements is a unit vector that has all zero entries except one location. For example, for $d=2$, these constitute up-right paths on $\mathbb{Z}^2$. Let $\Pi({\bf z})$ be the set of all directed paths from the origin to ${\bf z}$. The {\it last passage time} from the origin to ${\bf z}$ is defined as
\begin{equation}
T(\bz) = \max_{\pi \in \Pi({\bf z})} \sum_{\bv \in \pi}  X({\bf z})\;,
\end{equation}
which is the random variable given by the weight of the directed path $\pi$ from the origin to $\bz$  with the maximum sum. Despite the terminology ``last passage time", we allow the random variables $X({\bf z})$ to be negative. The terminology is a carry-over from first-passage percolation time where often the random variables are required to be positive, even though that will not be necessary herein.

We will be interested in the function
\begin{equation}
g(\bx) = \sup_{n \in \mathbb{N}} \frac{1}{n} E[T(\lfloor {n\bx} \rfloor )]
\end{equation}
where the floor function $\lfloor \cdot \rfloor$ is interpreted elementwise. From \cite{martin} we have the following properties for $g(\bx)$.
\begin{itemize}
\item $\lim_{n \rightarrow \infty} \frac{1}{n} T(\lfloor {n\bx} \rfloor ) = g(\bx)$, almost surely.
\item $g(\alpha \bx)=\alpha g(\bx)$ for $\alpha >0$
\item $g(\bx)$ is a symmetric function
\item $g(\bx)$ is super-additive: $g(\bx) + g(\by) \leq g(\bx+\by)$ for $\bx,\by \in \mathbb{R}_+^d$
\item $g(\bx)$ is concave on $\mathbb{R}_+^d$
\end{itemize}
\subsection{Schur-Concavity of $g(\bx)$}
Since $g(\bx)$ is homogeneous and $\bx \in \mathbb{R}_+^d$ with nonnegative entries, to fully define $g(\bx)$ it
sufficies to consider $\bx$ on the probability simplex ${\cal P} := \{{\bf p} \in \mathbb{R}_+^d: \parallel {\bf p} \parallel_1 =1\}$. Let $\be_i$ be the $d$-dimensional unit vector in the $i^{th}$ direction. Since $\Pi(\be_i)$ only contains a single path, $T(\be_i)$ does not involve a maximization, and $g(\be_i)=\mu$ due to the law of large numbers. Hence, when $\bx$ is on the probability simplex, $g(\bx)$ is minimized if $\bx=\be_i$, a unit vector concentrated along the $i^{th}$ dimension.

In fact, since $g(\bx)$ is concave and symmetric, it is also Schur-concave \cite{marshall-olkin}. Loosely speaking, this means that $g(\bx)$  is smaller on the probability simplex when $\bx$ is concentrated along a few directions only. To state this more precisely, we recall the definition of majorization. Let $x_{[i]}$ be a sorting of the elements of $\bx$ so that $x_{[1]} \geq x_{[2]} \ldots \geq  x_{[d]}$.  We denote the majorization of $\bx$ by $\by$ as $\bx  \prec \by$  defined as
\begin{equation}
\sum_{i=1}^k x_{[i]} \leq \sum_{i=1}^k y_{[i]}  \;\;\;\; k=1,\ldots, d-1 \;, \label{maj}
\end{equation}
and equality holds in (\ref{maj}) when $k=d$. Schur-concavity of $g(\bx)$ means that $\bx  \prec \by$ implies $g(\by) \leq g(\bx)$.
\subsection{Upperbound on $g(\bx)$}
We now derive a large-deviation-based upper bound on $g(\bx)$ which depends on the entropy of the normalized direction vector defined by $\bx$. This bound was derived in \cite{whitt} for $d=2$, where the connection with entropy was not provided.

Since the entries of $\bx$ are nonnegative, and $g(\bx)$ is homogeneous, we will focus on upperbounding $g(\bp)$ where $\bp$ is a probability vector that will have the interpretation of the empirical distribution of the fraction of total number of steps taken in each dimension. The upperbound for $g(\bx)$ will require the large-deviation rate function of the random variable $X({\bf z})$ defined as
\begin{equation}
I(x) := \sup_{\nu >0} \left(x \nu - \log(M(\nu)) \right)\;,
\end{equation}
where $M(\nu) := E[\exp(\nu X(\bz)] $. Note that $I(\cdot)$ is monotonically increasing for arguments greater than $\mu$. The upperbound is now given in the following theorem:
\\
{\bf Theorem:} Let $\bp \in \cal{P}$ be a probability vector with rational entries, and assume $E[\exp(\nu X(\bz)]$ exists for some $\nu$ in the neighborhood of the origin. Then
\begin{equation}
g(\bp) \leq I_{+}^{-1}(H(\bp))
\label{bound}
\end{equation}
where
\begin{equation}
I_{+}^{-1}(\beta) := \inf_{\theta} \{\theta: \beta < I(\theta), {\rm and} \;\; \theta>\mu \} \;,
\end{equation}
is the inverse of the monotonic portion of the rate function above the mean,
and $H(\bp)$ is the entropy function of a $d$ dimensional probability vector \cite{cover}.\\
{\bf Proof:} We will show that
\begin{eqnarray}
\lim_{l \rightarrow \infty} P\left[\frac{1}{l} T(l \bp) > \alpha \right] = 0
\label{astx}
\end{eqnarray}
exponentially, whenever $\alpha > I_{+}^{-1}(H(\bp))$.

Since $\bp$ has rational entries, there exists an integer $m$ such that $\bx := m \bp \in
\mathbb{Z}^d$,
so that
$\parallel \bx \parallel_1 =m$.
Setting $l=mn$ we can write
\begin{equation}
\frac{1}{l} T(l\bp) = \frac{1}{mn} T(\bx n) = \max_{\pi \in \Pi(n\bx)}
\frac{1}{mn} \sum_{v \in \pi} X(\bv) \;.
\end{equation}
The number of paths in $\Pi(n \bx)$ can be counted and upperbounded using \cite[pp. 351]{cover}
\begin{equation}
|\Pi(n \bx)| = {nm \choose nmp_1 \ldots nmp_d} \leq \exp(nmH(\bp)) \;, \label{count} \\
\end{equation}
and each path $\pi$ in $\Pi(n\bx)$ involves $\parallel n\bx \parallel_1 = nm$ random variables. We have
\begin{eqnarray}
\label{zero}
P\left[\frac{1}{l} T(l \bp) > \alpha \right]    &=&
P\left[ \bigcup\limits_{\pi \in \Pi(n\bx)}
\left(
\frac{1}{mn} \sum_{v \in \pi} X(\bv)\right] > \alpha
\right)
\\
\label{one}
&\leq&
\exp(nmH(\bp)) P\left[
\left(
\frac{1}{mn} \sum_{v \in \pi} X(\bv)\right) > \alpha
\right]  \\ \label{two} &\leq &
\exp(nmH(\bp)) \exp(-nm I(\alpha))
 \\ \label{three} &\leq &
\exp(-l(I(\alpha)-H(\bp)))
\end{eqnarray}
where (\ref{zero}) can be written as a union due to the max operator, (\ref{one})
is due to the union bound and (\ref{count}), (\ref{two}) is due to the Chernoff bound, and (\ref{three}) is obtained by substituting $l=nm$, and the monotonicity
of $I(\cdot)$ for arguments above $\mu$. This proves that the probability in question decays exponentially in $l$, whenever $\alpha > I_{+}^{-1}(H(\bp))$.

Note that if one is interested in bounding $g(\bx)$ with $\parallel \bx \parallel_1 \neq 1$, one can use the homogeneity of $g(\bx)$. If $\bp$ has irrational entries, a rational approximation along with continuity arguments will show that the bound in the Theorem is still valid. It is well-known that the entropy function is Schur-concave in the probability vector. Due to the monotonically increasing nature of the rate function and its inverse, $I_{+}^{-1}(H(\bp))$ is also Schur-concave, just like the quantity $g(\bp)$ it is bounding in (\ref{bound}).
Hence, one appealing feature of the upperbound is that it retains the Schur-concavity of the quantity it is bounding.
\bibliographystyle{unsrt}
\bibliography{lpp_it}
\end{document}